# Templated synthesis of cyclic poly(ionic liquid)s

Qingquan Tang,[a] Weiyi Zhang,[b] Jiayin Yuan[b] and Qiang Zhao[a]*

a. Key Laboratory of Material Chemistry for Energy Conversion and Storage, Ministry of Education, School of Chemistry and Chemical Engineering, Huazhong University of Science and Technology, Wuhan 430074 China.
b. Department of Materials and Environmental Chemistry, Stockholm University, Stockholm 10691 Sweden.

* Corresponding author, E-mail: zhaoq@hust.edu.cn.

**Abstract**

Charged cyclic polymers, e.g. cyclic DNAs and polypeptides, play enabling roles in organisms, but their synthesis was challenging due to the well-known "polyelectrolyte effect". To tackle the challenge, we developed a templated method to synthesize a library of imidazolium and pyridinium based cyclic poly(ionic liquid)s. Cyclic templates, cyclic polyimidazole and poly(2-pyridine), were synthesized first through ring-closure method by light-induced Diels−Alder click reaction. Through quaternization of cyclic templates followed by anion metathesis, the cyclic poly(ionic liquid)s were synthesized, which paired with varied counter anions.

# 1. Introduction

Cyclic polymers (CPs) feature unique properties compared to their linear counterparts [1-4], such as lower melt viscosity, smaller hydrodynamic volume and modulated crystallization kinetics, leading to functional materials ranging from gels [5], micelles [6], to nanotubes [7], etc. CPs are normally synthesized through ring-expansion and/or ring-closure methods [8-10]. In contrast to the vast majority of synthetic CPs that are neutral, biology is replete with natural cyclic polymers bearing charge, such as cyclic deoxyribonucleic acid [11-13] and circular polypeptides [14-16], which were discovered in virus and bacterial [17,18], etc. The combination of charge and cyclic topology is crucial for properties such as self-replication [13], biological recognition [11] and activities [14]. As such, charged cyclic polymers, i.e., cyclic polyelectrolytes, are emerging topics attracting growing interest from both the chemistry and materials perspectives. For example, cyclic cationic poly((2-dimethylamino) ethylmethacrylate) exhibited reduced cytotoxicity than its linear counterparts in gene transfer processes [19]. In this context, we are particularly interested in cyclic poly(ionic liquid)s (CPILs), given that PILs is a subclass of ionic polymers highlighting beneficial properties [20-24] including surface activity, adaptive solubility, ionic conductivity, wide electrochemical windows, etc. As an emerging family of functional polymers, PILs have been exploited as membranes [25-28], self-assembly [29-31], ion conductors [32,33], structured carbon [34,35], advanced catalysis [36-38], etc. In this regard, CPILs can exhibit some novel properties than the linear counterparts, such as smaller hydrodynamic volume, unique diffusion dynamic [39], and higher biological activity [19]. As such, CPILs are potentially applicable for self-assembled entities with unique rheological properties, improved bioactivities, and advanced conducting abilities.

Different from neutral CPs synthesized by ring-closure cyclization at extremely diluted polymer solutions [40], synthesis of CPILs suffers from the so-called "polyelectrolyte effect" [41,42]. Under this circumstance, the electrostatic repulsion of charged chains renders collision of chain ends in dilute solution particularly difficult [43,44], thus impeding effective ring-closure. The ring-closure cyclization efficiency could be improved by increasing the polymer solution concentration, at the price of the enhanced intermolecular reaction and polycondensation byproducts. As an alternative approach, the ring-expansion method is more suited for monomers of specific chemical structures, such as strain olefin [45,46] and lactones [47,48], precluding the majority of ionic liquid monomers such as the most extensively studied the imidazolium ones. Putting together, the synthesis of CPILs is an attractive topic but never reported so far.

Herein, we developed a templated method to synthesize CPILs (**Scheme 1**). Neutral cyclic polymer templates were synthesized first, which were subsequently quaternized and ion exchanged to yield target CPILs. This "template first" strategy skipped the adverse "polyelectrolyte effect" involved in the ring-closure method. In detail, linear poly(2-vinylpyridine) (LP2VP) and poly(1-(4-vinylbenzyl)imidazole) (LPVBIm) were synthesized by Reversible Addition-Fragmentation Chain Transfer (RAFT) polymerization, followed through cyclization reaction by Diels-Alder click reaction of orthoquinodimethane and dithioester chain ends under UV light (365 nm) [49]. Thereafter, cyclic poly(1-cyanomethyl-2-vinylpyridine bromine) (CP2VP-Br) and cyclic poly(1-cyanomethyl-(4-vinylbenzyl)imidazole bromine) (CPVBIm-Br), were synthesized by quaternization of the cyclic polymer precursors with bromoacetonitrile (BrCH$_2$CN). Subsequently, a library of CPILs paired with different counter anions was obtained by counter-ion exchange. They are termed CP2VP-X and CPVBIm-X, where X is tetrafluoroborate (BF$_4$), hexafluorophosphate (PF$_6$), or bis(trifluoro methanesulfonyl)imide) (Tf$_2$N).

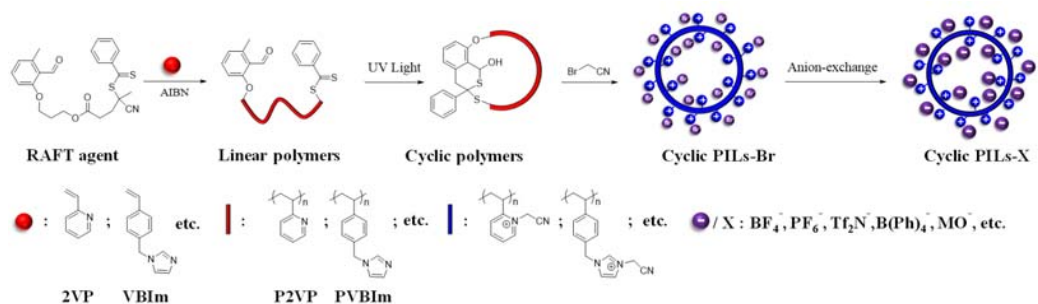

**Scheme 1** Templated synthesis of CPILs.

## 2. Experimental

*2.1 Materials*

Glyoxaline, 4-(chloromethyl)styrene, bromoacetonitrile, sodium tetrafluoroborate (NaBF$_4$), potassium hexafluorophosphate (KPF$_6$), lithium bis(trifluoro-methanesulfonyl)imide (LiTf$_2$N), sodium tetraphenylborate (NaB(Ph)$_4$) and methyl orange (MO), sodium hydroxide, imidazole, 4-vinylbenzyl chloride，were purchased as reagent grade from Aldrich, Acros, Alfa Aesar, Aladdin, and used as received. Qetroleum ether, methanol, diethyl ether, acetonitrile, dichloromethane (DCM), tetrahydrofuran (THF), hexane, chloroform ($CHCl_3$), *N,N*-Dimethylformamide (DMF) were purchased as regent grade from Beijing Chemical Reagent Co. and used as received unless otherwise noted. 2-vinyl pyridine (2VP) were dried over $CaH_2$ and distilled before use. 2,2'-Azoisobutyronitrile (AIBN) was recrystallized from ethanol and stored at 4 °C. RAFT agent (3-(2-formyl-3-methylphenoxy)propyl 4-cyano-4-((phenylcarbonothioyl)thio)pentanoate) [49], and *N*-(4-vinylbenzyl)-imidazole (VBIm) [50], were synthesized according to the previous literatures. A low-pressure mercury lamp (120 W) (CEL-LPH120-254, Beijing China Education Au-light co. Ltd) was used as the UV light source.

*2.2 Characterization*

$^1$H-NMR spectrum were recorded on a Bruker Avance 400 spectrometer at room temperature. Ultraviolet Spectrum were recorded using a TU-1901 Ultraviolet Spectrophotometer. FT-IR spectrum were recorded on a Thermo Nicolet iS5 Spectrometer at room temperature. Gel permeation chromatography (GPC) in DMF was conducted on a system comprised of a Waters 515 HPLC pump, and a Waters 2414 RI detector equipped with four Waters Styragel columns (HT 2, HT 3, HT 4, and HT 5). DMF with 0.01 M LiBr was used as the eluent at a flow rate of 1.0 ml/min. Polystyrene standards were used for the calibration. Matrix-assisted laser desorption and ionization time-of-flight (MALDI-TOF) mass spectrometry was performed on a Bruker Biflex III spectrometer equipped with a 337 nm nitrogen laser. Inductively coupled plasma mass spectrometry (ICP-MS) were recorded on a PerkinElmer NexION 300X. Elemental analysis (C, H, N, S) was performed on a Flash EA1112 from Thermo Quest Italia S.P.A.

*2.3 Synthesis of LP2VP*

RAFT agent (45.5 mg, 0.1 mmol) and AIBN (3.6 g 0.22 mmol) were dissolved in 2VP (3.15 g, 30 mmol) on stirring under inert atmosphere. The clear solution was degassed via three freeze-thaw-pump cycles. After stirring at 60 °C for 7.5 h, the reaction was terminated by exposure to air. Polymer was precipitated from DCM solvent into the excess of hexane three times. After drying overnight in a vacuum oven at room temperature, the light red powder was obtained with a monomer conversion of 15.2% from $^1$H-NMR with a yield of 83.2%.

*2.4 Synthesis of LPVBIm*

RAFT agent (45.5 mg, 0.1 mmol) and AIBN (3.28 g 0.2 mmol) were dissolved in VBIm (1.84 g, 10 mmol) on stirring under inert atmosphere. The clear solution was degassed via three freeze-thaw-pump cycles. After stirring at 60 °C for 17 h, the reaction was terminated by exposure to air. Polymer was precipitated from DCM solvent into the excess of diethyl ether three times. After drying overnight in a vacuum oven at room temperature, the light red powder was obtained with a monomer conversion of 32.2% from $^1$H-NMR with a yield of 90.2%.

*2.5 Synthesis of cyclic poly(2-vinylpyridine) (CP2VP)*

After dissolving linear precursors (50 mg) in acetonitrile (1000 ml), the solution was stirred under UV light irradiation for 9 h at room temperature. Pure cyclic polymers were conveniently collected by evaporation of solvent. Repeat the procedure 5 times until enough CP2VP were obtained for quaternization.

*2.6 Synthesis of cyclic poly(1-(4-vinylbenzyl)imidazole) (CPVBIm)*

Follow similar procedures in "synthesis of CP2VP", but with less linear precursors (20 mg) in acetonitrile (1000 ml).

*2.7 Synthesis of CP2VP-Br*

A solution of CP2VP (200 mg) in bromoacetonitrile (5 ml) was stirred at 60 °C for 3 days. After the reaction, the solution was precipitated in diethyl ether (50 ml), and then the precipitations were collected. The precipitations were re-dissolved in water (1 ml), and precipitated in THF (20 ml) again. After 3 dissolving-precipitations circles, the CP2VP-Br was obtained upon vacuum oven at room temperature overnight.

*2.8 Synthesis of CPVBIm-Br*

A mixture solution of CPVBIm (200 mg) with bromoacetonitrile (3 ml) in NMP (2

ml) was stirred at 60 °C for 1 days. After the reaction, the solution was dropped into diethyl ether (50 ml), and then precipitations were collected. The precipitations were re-dissolved in water (1 ml), and precipitated in THF (20 ml) again. After 3 dissolving-precipitations circles, the CPVBIm-Br was obtained upon vacuum oven at room temperature overnight.

*2.9 Synthesis of Various Cyclic PILs*

Ten different cyclic PILs were prepared following identical counter-ion exchange procedures, with molar ratios of 2:1 (salts : PILs). The dissolved PILs solution (*e.g.* 50 mg CP2VP-Br/CPVBIm-Br in 2 ml water) were dropped into salt solution (*e.g.* 36 mg $NaBF_4$ in 2ml water), and after filtration, the precipitations were washed 3 times with water, and dried in vacuum oven at room temperature overnight for next steps.

*2.10 Calculation of counter-ion exchange efficiency*

The efficiency of counter-ion exchange (E%) is defined as the ratio between exchanged ionic groups and ionic pyridine or imidazole units (equation 1 below).

$$E\% = \frac{exchanged\ ionic\ groups}{ionic\ pyridine\ or\ imidazole\ units} \times 100\% \qquad (1)$$

**3. Results and discussions**

*3.1 Synthesis and characterization of cyclic templates*

GPC characterization of LP2VP (**Fig. 1A black**) reveal a well-defined, monomodal, and symmetric elution trace, indicative of the successful RAFT polymerization with fine control over molecular weight distribution. The whole GPC trace of CP2VP (**Fig. 1A red**) remains the same to LP2VP but shifts to a lower molecular weight regime. It indicates a smaller hydrodynamic radius of CP2VP and the corresponding apparent molecular weight ($M_n$=11490, PDI=1.15) in comparison to LP2VP precursor ($M_n$=14020, PDI=1.12). MALDI-TOF mass spectrum of LP2VP (**Fig. 2A**) and CP2VP (**Fig. 2B**) indicates that the absolute molecular weights are similar for both cases expanding from 4500 to 7500 and centering at 6000. Combining the much smaller apparent $M_n$ of CP2VP than that of linear precursor from GPC, the successful cyclization is demonstrated [51-53]. Chemical structures of CP2VP were characterized by UV-Vis spectra (**Fig. 3**) and $^1$H NMR (**Fig. 4**), whereas the π−π* adsorption peak of the thiocarbonyl moiety at 305 nm disappears after ring closure reaction (**Fig. 3A** red arrow). Additionally, the NMR signal of orthoquinodimethane end group in LP2VP (at 10.66 ppm, $H_a$ in Fig. 4A) vanishes after Diels-Alder ring-closure reaction (Fig. 4B), in good agreement with the UV-Vis results.

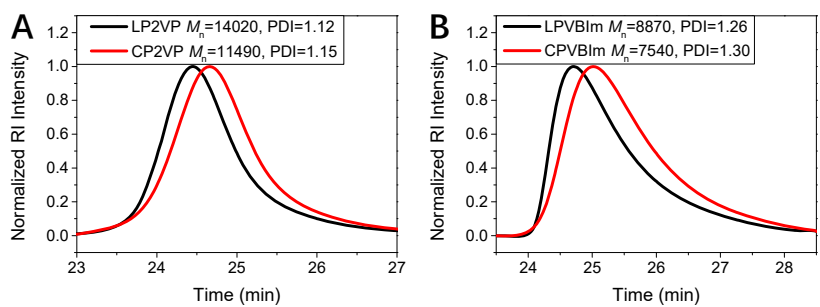

**Fig. 1** A) GPC traces of (A) LP2VP and CP2VP, and (B) LPVBIm and CPVBIm. Eluent: DMF; calibration standard: polystyrene.

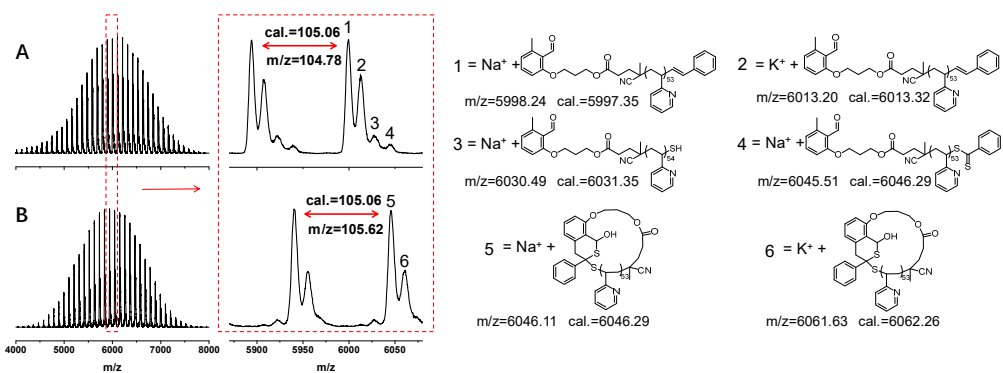

**Fig 2.** MALDI−TOF mass spectrum for LP2VP precursor and the corresponding CP2VP.

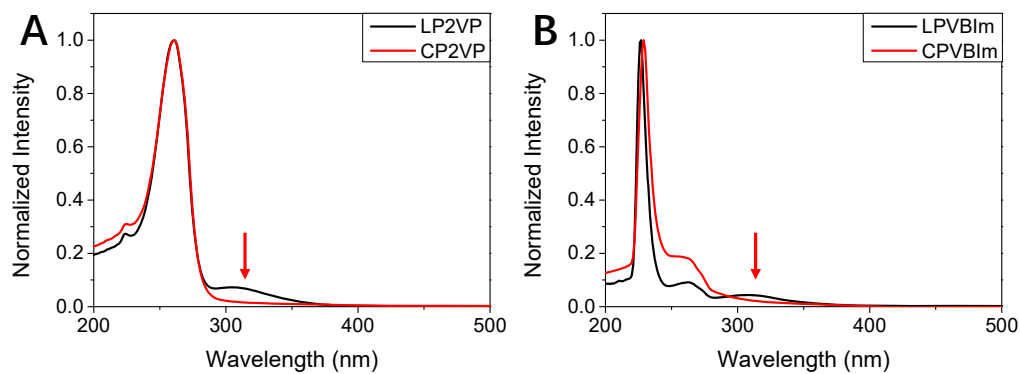

**Fig 3.** UV-vis spectrum of (A) LP2VP (black) and the resultant CP2VP (red) in DCM, (B) LPVBIm (black) and the resultant CPVBIm (red) in DCM.

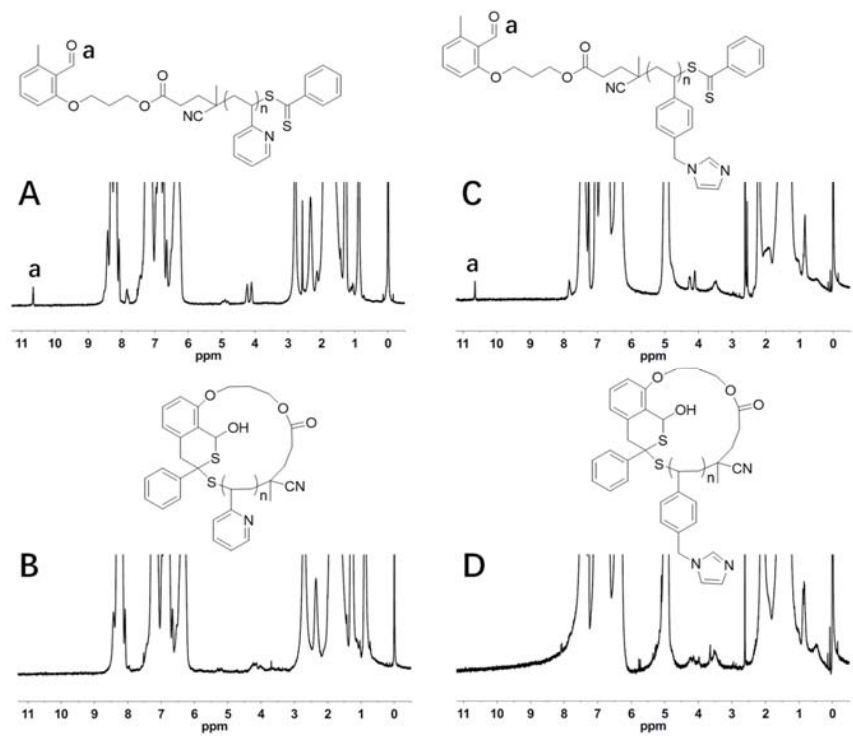

**Fig 4.** $^1$H NMR spectrum of LP2VP (A), CP2VP (B), LPVBIm (C) and CPVBIm (D) in CDCl$_3$.

Similar results were also found for the synthesis of cyclic PVBIm (CPVBIm) from its linear counterpart LPVBIm. The GPC traces (**Fig. 1B**) in a well-defined, monomodal, and symmetrical shape are observed for LPVBIm ($M_n$=8870, PDI=1.26) and CPVBIm ($M_n$=7540, PDI=1.30), respectively. Moreover, the disappearance of orthoquinodimethane and thiocarbonyl group as a result of Diels-Alder ring closure reaction was confirmed by $^1$H NMR (**Fig. 4D**) and UV-vis (**Fig. 3B**), respectively. These results all support the successful synthesis of CP2VP and CPVBIm.

*3.2 Synthesis and characterization of cyclic PILs*

The two targeted CPILs, *i.e.* CP2VP-Br and CPVBIm-Br, were synthesized by quaternization of CP2VP and CPVBIm with BrCH$_2$CN, which is an active quaternizing agent with beneficial hydrophilicity for improving anion metathesis efficiency. In their $^1$H NMR spectrum, new signals of H$_b$ (5.90 ppm, **Fig. 5B**) and H$_c$ (5.50 ppm, **Fig. 5D**) appear for CP2VP-Br and CPVBIm-Br. Both signals can be assigned to the newly formed -C*H*$_2$-CN groups, as a result of the quaternization of CP2VP and CPVBIm by BrCH$_2$CN. Through the area integration ratio between H$_b$ (-C*H*$_2$-CN) and H$_a$ (pyridine units), the quaternization degree of CP2VP is calculated to be ca. 60%. In a similar way, the quaternization degree of CPVBIm-Br is *ca.* 95%, i.e. being more efficient than CP2VP. From FT-IR spectrum (**Fig. 6**), two new absorption bands (dashed rectangles) are seen at 1200 cm$^{-1}$ and 2200 cm$^{-1}$, which are assigned to the vibration mode of newly formed -C*H*$_2$-CN, in good agreement with the NMR results (**Fig. 5**).

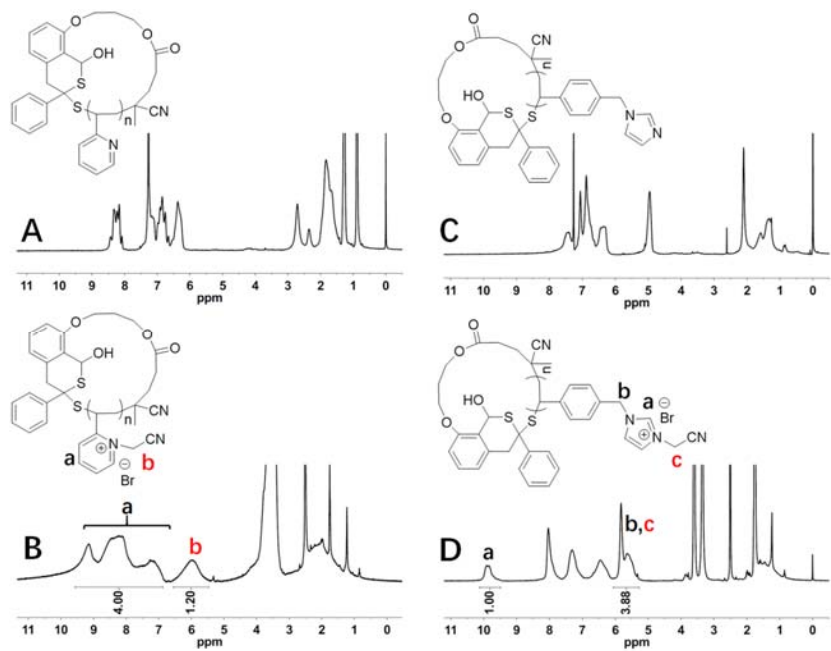

**Fig. 5** $^1$H NMR spectrum of A) CP2VP and C) CPVBIm in CDCl$_3$; B) CP2VP-Br and D) CPVBIm-Br in DMSO-*d$_6$*.

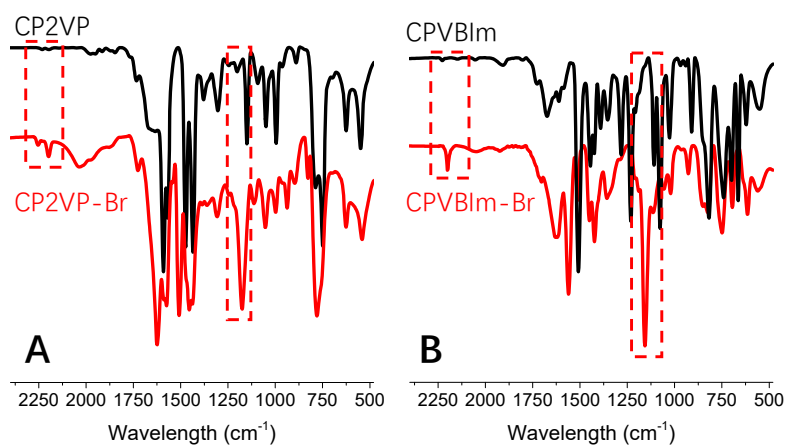

**Fig. 6** FT-IR spectrum of CP2VP (A, in black), CP2VP-Br (A, in red), CPVBIm (B, in black) and CPVBIm-Br (B, in red).

*3.3 Synthesis and characterization of various cyclic PILs*

Counter-anion exchange was conducted to expand the structural spectrum of the as-synthesized CPILs. The successful synthesis of CP2VP- B(Ph)$_4$/MO and CPVBIm-B(Ph)$_4$/MO (MO ~ methyl orange) were verified by $^1$H NMR spectrum (**Fig. 7**), whereas the exchange efficiency was also calculated. For both CP2VP-X (**Fig. 8A**) and CPVBIm-X (**Fig. 8B**), characteristic bands of each counter anions (BF$_4^-$, PF$_6^-$, Tf$_2$N$^-$, B(Ph)$_4^-$, MO$^-$) are clearly seen from FT-IR curves of the corresponding CPIL-X after counter-ion exchange (dashed rectangles, **Fig. 8**). The efficiency of counter-ion exchange, defined as the ratio of exchanged repeating ionic units to their overall repeating ionic unit, was further quantified (**Fig. 9**). The exchange efficiency from Br$^-$ to Tf$_2$N$^-$ is 71.3% and 76.2% for CP2VP-X and CPVBIm-X respectively, and the exchange efficiency of B(Ph)$_4^-$ for both polymers is less than 85%. Notably, the anion exchange degree of CP2VP-Br or CPVBIm-Br with MO anion are 99.1% and 98.1%, respectively, and this quantitative exchange is in agreement with previous works [54,55]. Moreover, the ion exchange degree of CP2VP-Br and CPVBIm-Br with BF$_4^-$ are 81.4% and 93.4%, respectively, as determined by ICP-MS.

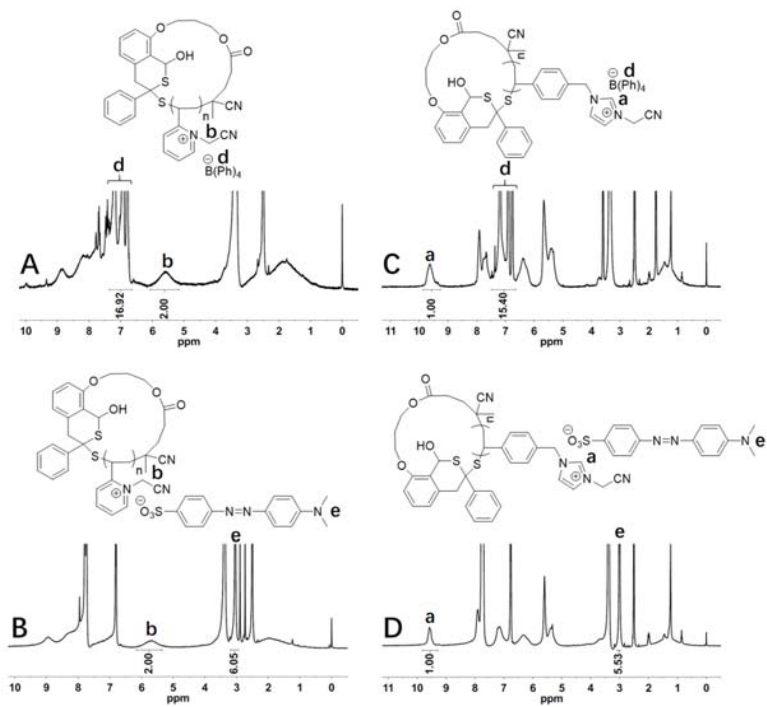

**Fig 7.** $^1$H NMR spectrum of CPILs: A) CP2VP-B(Ph)$_4$; B) CP2VP-MO; C) CPVBIm-B(Ph)$_4$; D) CPVBIm-MO; in DMSO-*d$_6$*.

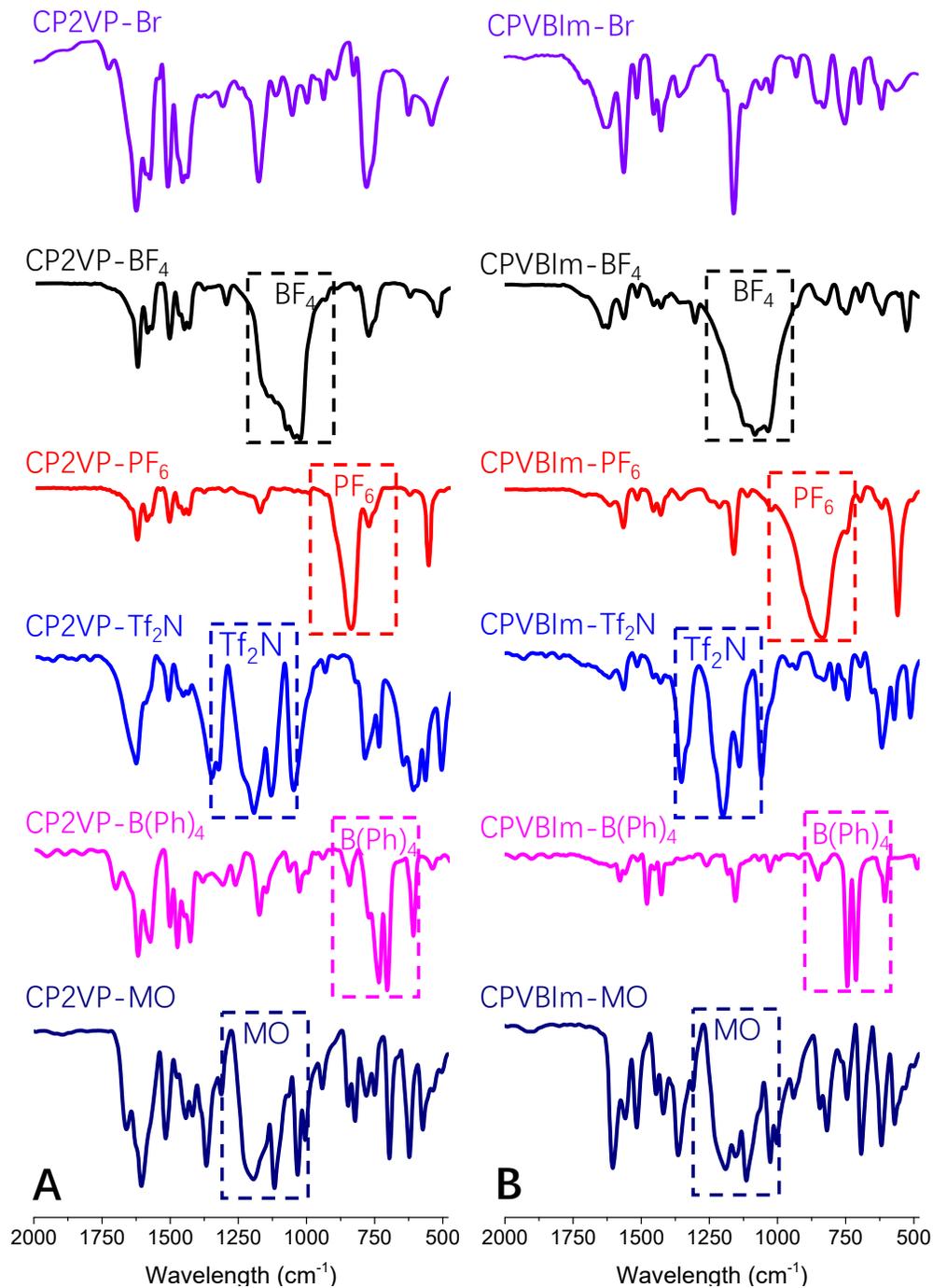

**Fig. 8** FT-IR spectrum of CP2VP-Br (A, in purple), CP2VP-BF4 (A, in black), CP2VP-PF6 (A. in red), CP2VP-Tf2N (A. in blue), CP2VP-B(Ph)4 (A. in red purple), CP2VP-MO (A, in navy), CPVBIm-Br (B, in purple), CPVBIm-BF4 (B, in black), CPVBIm-PF6 (B, in red), CPVBIm-Tf2N (B, in blue), CPVBIm-B(Ph)4 (B, in red purple), CPVBIm-MO (B, in navy blue).

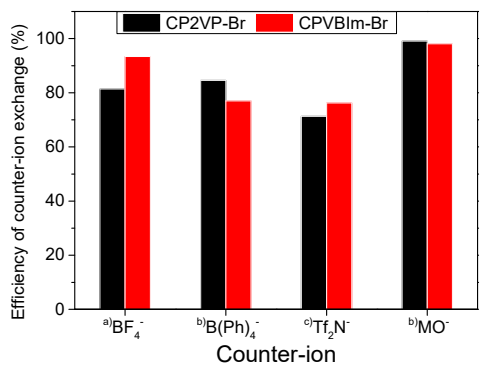

**Fig. 9** Efficiency of counter-ion exchange of CP2VP-Br and CPVBIm-Br with varied anions ($BF_4^-$, $Tf_2N^-$, $B(Ph)_4^-$, $MO^-$). a) determined from ICP-MS. b) determined from NMR. c) determined from Element analysis.

## 4. Conclusions

A templated method was proposed to synthesize pyridinium and imidazolium based cyclic PILs paired with various counter anions. LP2VP and LPVBIm were synthesized by RAFT polymerization, and the ring-closure was accomplished with quantitation through the UV induced Diels-Alder reaction. Quaternization of CP2VP and CPVBIm yielded CPILs whose counter anions could be effectively exchanged with other organic anions such as $BF_4^-$, $PF_6^-$, $Tf_2N^-$, $B(Ph)_4^-$ and $MO^-$. Given that both the cations (pyridinium and imidazolium) and anions are most frequently used for ionic liquids and PILs, this work paves the synthetic path to a broad spectrum of CPILs with enriched topology, self-assembly and implication potentials, which have not been addressed due to the lack of suitable synthetic tools.

**Conflicts of interest**

There are no conflicts to declare.

**Acknowledgements**

Q. Z. gratefully acknowledges financial support from the Huazhong University of Science and Technology (No. 3004013118). Q. T. acknowledges financial support from the Selected Postdoctoral Technology Foundation in Huibei Provinece of China. J. Y. acknowledges financial support from the European

Research Council (ERC) Starting Grant (No. 639720-NAPOLI), the Wallenberg Academy Fellow program (No. KAW2017.0166) from the Knut and Alice Wallenbergs Foundation in Sweden, and the Strategic Fund from Stockholm University (Contract number: SU FV-2.1.1-005).

**Notes and references**

[1] X. Y. Tu, M.-Z. Liu and H. Wei, Recent progress on cyclic polymers: synthesis, bioproperties, and biomedical applications, J. Polym. Sci., Part A: Polym. Chem. 54(2016) 1447-1458.

[2] E. J. Shin, W. Jeong, H. A. Brown, B. J. Koo, J. L. Hedrick and R. M. Waymouth, Crystallization of cyclic polymers: Synthesis and crystallization behavior of high molecular weight cyclic poly($\varepsilon$-caprolactone)s, Macromolecules 44(2011) 2773-2779.

[3] T. Yamamoto, Synthesis of cyclic polymers and topology effects on their diffusion and thermal properties, Polym. J. 45(2012) 711-717.

[4] T. Yamamoto and Y. Tezuka, Cyclic polymers revealing topology effects upon self-assemblies, dynamics and responses, Soft Matter 11(2015) 7458-7468.

[5] K. Zhang, M. A. Lackey, J. Cui and G. N. Tew, Gels based on cyclic polymers, J. Am. Chem. Soc. 133(2011) 4140-4148.

[6] B. Zhang, H. Zhang, Y. Li, J. N. Hoskins and S. M. Grayson, Exploring the effect of amphiphilic polymer architecture: Synthesis, characterization, and self-assembly of both cyclic and linear poly(ethylene gylcol)-b-polycaprolactone, ACS Macro Lett. 2(2013) 845-848.

[7] M. Schappacher and A. Deffieux, Synthesis of macrocyclic copolymer brushes and their self-assembly into supramolecular tubes, Science 319(2008) 1512-1515.

[8] T. Josse, J. De Winter, P. Gerbaux and O. Coulembier, Cyclic polymers by ring-closure strategies, Angew. Chem. Int. Ed. 55(2016) 13944-13958.

[9] B. A. Laurent and S. M. Grayson, Synthetic approaches for the preparation of cyclic polymers, Chem. Soc. Rev. 38(2009) 2202-2213.

[10] Z. Jia and M. J. Monteiro, Cyclic polymers: Methods and strategies, J. Polym. Sci., Part A: Polym. Chem. 50(2012) 2085-2097.

[11] E. T. Kool, Recognition of DNA, RNA, and proteins by circular oligonucleotides, Acc. Chem. Res. 31(1998) 502-510.

[12] A. H. El-Sagheer, R. Kumar, S. Findlow, J. M. Werner, A. N. Lane and T. Brown, A very stable cyclic DNA miniduplex with just two base pairs, Chembiochem 9(2008) 50-52.

[13] J. Kim, J. Lee, S. Hamada, S. Murata and S. H. Park, Self-replication of DNA rings, Nat. Nanotech. 10(2015) 528-534.

[14] K. W. Plaxco, K. T. Simons, I. Ruczinski and B. David, Topology, stability, sequence, and length: Defining the determinants of two-state protein folding kinetics, Biochemistry 39(2000) 11177-11183.

[15] J. A. Camarero, D. Fushman, S. Sato, I. Giriat, D. Cowburn, D. P. Raleigh and T.

W. Muir, Rescuing a destabilized protein fold through backbone cyclization, J. Mol. Biol. 308(2001) 1045-1062.

[16] J. A. Camarero and T. W. Muir, Biosynthesis of a head-to-tail cyclized protein with improved biological activity, J. Am. Chem. Soc. 121(1999) 5597-5598.

[17] D. R. Helinski and D. B. Clewell, Circular DNA, Annu. Rev. Biochem. 40(1971) 899-942.

[18] R. Borra and J. A. Camarero, Recombinant expression of backbone-cyclized polypeptides, biopolymers 100(2013) 502-509.

[19] H. Wei, D. S. H. Chu, J. Zhao, J. A. Pahang and S. H. Pun, Synthesis and evaluation of cyclic cationic polymers for nucleic acid delivery, ACS Macro Lett. 2(2013) 1047-1050.

[20] J. Yuan, D. Mecerreyes and M. Antonietti, Poly(ionic liquid)s: An update, Prog. Polym. Sci. 38(2013) 1009-1036.

[21] D. Mecerreyes, Polymeric ionic liquids: Broadening the properties and applications of polyelectrolytes, Prog. Polym. Sci. 36(2011) 1629-1648.

[22] J. Lu, F. Yan and J. Texter, Advanced applications of ionic liquids in polymer science, Chem. Soc. Rev. 34(2009) 431-448.

[23] W. Zhang, Q. Zhao and J. Yuan, Porous polyelectrolytes: The interplay of charge and pores for new functionalities, Angew. Chem. Int. Ed. 57(2018) 6754-6773.

[24] N. Patil, D. Cordella, A. Aqil, A. Debuigne, S. Admassie, C. Jérôme and C. Detrembleur, Surface- and redox-active multifunctional polyphenol-derived poly(ionic liquid)s: Controlled synthesis and characterization, Polymer 49(2016) 7676-7691.

[25] Q. Zhao, J. W. C. Dunlop, X. Qiu, F. Huang, Z. Zhang, J. Heyda, J. Dzubiella, M. Antonietti and J. Yuan, An instant multi-responsive porous polymer actuator driven by solvent molecule sorption, Nat. Commun. 5(2014) 4293-4301.

[26] Q. Zhao, J. Heyda, J. Dzubiella, K. Taeuber, J. W. C. Dunlop and J. Yuan, Sensing solvents with ultrasensitive porous poly(ionic liquid) actuators, Adv. Mater. 27(2015) 2913-2917.

[27] C. C. Ye, F. Y. Zhao, J. K. Wu, X. D. Weng, P. Y. Zheng, Y. F. Mi, Q. F. An and C. J. Gao, Sulfated polyelectrolyte complex nanoparticles structured nanoflitration membrane for dye desalination, Desalination 307(2017) 526-536.

[28] R. Ma, Y. L. Ji, Y. S. Guo, Y. F. Mi, Q. F. An and C. J. Gao, Fabrication of antifouling reverse osmosis membranes by incorporating zwitterionic colloids nanoparticles for brackish water desalination, Chem. Eng.J. 416(2017) 35-44.

[29] K. Manojkumar, D. Mecerreyes, D. Taton, Y. Gnanou and K. Vijayakrishna, Self-assembly of poly(ionic liquid) (PIL)-based amphiphilic homopolymers into vesicles and supramolecular structures with dyes and silver nanoparticles, Polym. Chem. 8(2017) 3497-3503.

[30] A. J. Erwin, W. Xu, H. He, K. Matyjaszewski and V. V. Tsukruk, Linear and star poly(ionic liquid) assemblies: Surface monolayers and multilayers, Langmuir 33(2017) 3187-3199.

[31] D. Cordella, F. Ouhib, A. Aqil, T. Defize, C. Jerome, A. Serghei, E. Drockenmuller, K. Aissou, D. Taton and C. Detrembleur, Fluorinated poly(ionic liquid) diblock

copolymers obtained by cobalt-mediated radical polymerization-induced self-assembly, ACS Macro Lett. 6(2017) 121-126.

[32] R. L. Weber, Y. Ye, A. L. Schmitt, S. M. Banik, Y. A. Elabd and M. K. Mahanthappa, Effect of nanoscale morphology on the conductivity of polymerized ionic liquid block copolymers, Macromolecules 44(2011) 5727-5735.

[33] P. Maksym, M. Tarnacka, A. Dzienia, K. Erfurt, A. Chrobok, A. Zięba, K. Wolnica, K. Kaminski and M. Paluch, A facile route to well-defined imidazolium-based poly(ionic liquid)s of enhanced conductivity via RAFT, Polym. Chem. 8(2017) 5433-5443.

[34] T. P. Fellinger, A. Thomas, J. Yuan and M. Antonietti, 25th anniversary article: "Cooking carbon with salt": Carbon materials and carbonaceous frameworks from ionic liquids and poly(ionic liquid)s, Adv. Mater. 25(2013) 5838-5854.

[35] J. Gong, H. Lin, M. Antonietti and J. Yuan, Nitrogen-doped porous carbon nanosheets derived from poly(ionic liquid)s: Hierarchical pore structures for efficient $CO_2$ capture and dye removal, J. Mater. Chem. A 4(2016) 7313-7321.

[36] Q. Zhao, P. Zhang, M. Antonietti and J. Yuan, Poly(ionic liquid) complex with spontaneous micro-/mesoporosity: Template-free synthesis and application as catalyst support, J. Am. Chem. Soc. 134(2012) 11852-11855.

[37] Y. Lu and M. Ballauff, Spherical polyelectrolyte brushes as nanoreactors for the generation of metallic and oxidic nanoparticles: Synthesis and application in catalysis, Prog. Polym. Sci. 59(2016) 86-104.

[38] J. Y. Yuan, S. Wunder, F. Warmuth and Y. Lu, Spherical polymer brushes with vinylimidazolium-type poly(ionic liquid) chains as support for metallic nanoparticles, Polymer 53(2012) 43-49.

[39] L. Liu, W. Chen and J. Chen, Shape and diffusion of circular polyelectrolytes in salt-free dilute solutions and comparison with linear polyelectrolytes, Macromolecules, 50(2017) 6659-6667.

[40] B. A. Laurent and S. M. Grayson, An efficient route to well-defined macrocyclic polymers via "Click" cyclization, J. Am. Chem. Soc. 128(2006) 4238-4239.

[41] A. R. Khokhlov and K. A. Khachaturian, On the theory of weakly charged polyelectrolytes, Polymer 23(1982) 1742-1750.

[42] B. A. V. Dobrynin, R. H. Colby and M. Rubinstein, Scaling theory of polyelectrolyte solutions, Macromolecules 28(1995) 1859-1871.

[43] M. F. Raymond and P. S. Ulrich, Electrostatic interaction of polyelectrolytes and simple electrolytes, J. Polym. Sci. 3(1948) 602-603.

[44] H. Yang, Q. Zheng and R. Cheng, New insight into "polyelectrolyte effect", Colloids and Surfaces A: Physicochem. Eng. Aspects., 2012, 407, 1-8.

[45] K. Zhang, M. A. Lackey, Y. Wu and G. N. Tew, Universal cyclic polymer templates, J. Am. Chem. Soc. 133(2011) 6906-6909.

[46] C. W. Bielawski, D. Benitez and R. H. Grubbs, An "endless" route to cyclic polymers, Science 297(2002) 2041-2044.

[47] M. Hong and E. Y. X. Chen, Completely recyclable biopolymers with linear and cyclic topologies via ring-opening polymerization of γ-butyrolactone, Nat. Chem. 8(2015) 42-49.

[48] S. Y. Hu, G. X. Dai, J. P. Zhao and G. Z. Zhang, Ring-opening alternating copolymerization of epoxides and dihydrocoumarin catalyzed by a phosphazene superbase, Macromolecules 49(2016) 4462-4472.

[49] Q. Tang, Y. Wu, P. Sun, Y. Chen and K. Zhang, Powerful ring-closure method for preparing varied cyclic polymers, Macromolecules 47(2014) 3775-3781.

[50] Y. Xie, Q. Sun, Y. Fu, L. Song, J. Liang, X. Xu, H. Wang, J. Li, S. Tu, X. Lu and J. Li, Sponge-like quaternary ammonium-based poly(ionic liquid)s for high $CO_2$ capture and efficient cycloaddition under mild conditions, J. Mater. Chem. A 5(2017) 25594-25600.

[51] T. Josse, O. Altintas, K. K. Oehlenschlaeger, P. Dubois, P. Gerbaux, O. Coulembier and C. Barner-Kowollik, Ambient temperature catalyst-free light-induced preparation of macrocyclic aliphatic polyesters, Chem. Commun. 50(2014) 2024-2026.

[52] P. Sun, Q. Tang, Z. Wang, Y. Zhao and K. Zhang, Cyclic polymers based on UV-induced strain promoted azide–alkyne cycloaddition reaction, Polym. Chem. 6(2015) 4096-4101.

[53] P. Sun, J. a. Liu, Z. Zhang and K. Zhang, Scalable preparation of cyclic polymers by the ring-closure method assisted by the continuous-flow technique, Polym. Chem. 7(2016) 2239-2244.

[54] S. Xiao, X. Lu and Q. Lu, Photosensitive polymer from ionic self-assembly of azobenzene dye and poly(ionic liquid) and its alignment characteristic toward liquid crystal molecules, Macromolecules 40(2007) 7944-7950.

[55] Q. Zhang, C. G. Bazuin and C. J. Barrett, Simple spacer-free dye-polyelectrolyte ionic complex: Side-chain liquid crystal order with high and stable photoinduced birefringence, Chem. Mater. 20(2008) 29-31.

# Supplementary Information

# Experimental

**Materials**

Glyoxaline, 4-(chloromethyl)styrene, bromoacetonitrile, sodium tetrafluoroborate (NaBF$_4$), potassium hexafluorophosphate (KPF$_6$), lithium bis(trifluoro-methanesulfonyl)imide (LiTf$_2$N), sodium tetraphenylborate (NaB(Ph)$_4$) and methyl orange (MO), sodium hydroxide, imidazole, 4-vinylbenzyl chloride，were purchased as reagent grade from Aldrich, Acros, Alfa Aesar, Aladdin, and used as received. Qetroleum ether, methanol, diethyl ether, acetonitrile, dichloromethane (DCM), tetrahydrofuran (THF), hexane, chloroform (CHCl$_3$), *N,N*-Dimethylformamide (DMF) were purchased as regent grade from Beijing Chemical Reagent Co. and used as received unless otherwise noted. 2-vinyl pyridine (2VP) were dried over CaH$_2$ and distilled before use. 2,2'-Azoisobutyronitrile (AIBN) was recrystallized from ethanol and stored at 4 °C. *N*-(4-vinylbenzyl)-imidazole[1], RAFT agent[2] were synthesized according to the previous literatures. A low-pressure mercury lamp (120 W) (CEL-LPH120-254, Beijing China Education Au-light co. Ltd) was used as the UV light source.

**Characterization**

$^1$H-NMR spectra were recorded on a Bruker Avance 400 spectrometer at room temperature.
Ultraviolet Spectra were recorded using a TU-1901 Ultraviolet Spectrophotometer.
FT-IR spectra were recorded on a Thermo Nicolet Avatar-330 Spectrometer at room temperature.
Gel permeation chromatography (GPC) in DMF was conducted on a system comprised of a Waters 515 HPLC pump, and a Waters 2414 RI detector equipped with four Waters Styragel columns (HT 2, HT 3, HT 4, and HT 5). DMF with 0.01 M LiBr was used as the eluent at a flow rate of 1.0 mL/min. Polystyrene standards were used for the calibration.

1. **Preparation of LP2VP**

A mixed solution of 2VP (3.15 g, 30 mmol), RAFT agent **1** (45.5 mg, 0.1 mmol) and AIBN (3.6 g 0.22 mmol) was degassed *via* three freeze-thaw-pump cycles. After stirring at 60 °C for 7.5 h, the reaction was terminated by exposure to air. Polymer was precipitated from an excess of hexane three times. After drying overnight in a vacuum oven at room temperature, the light red product was obtained with a monomer conversion of 15.2% from $^1$H-NMR.

## 2. Preparation of LPVBIm

A mixed solution of VBIm (1.84 g, 10 mmol), RAFT agent **1** (45.5 mg, 0.1 mmol) and AIBN (3.28 g 0.2 mmol) was degassed *via* three freeze-thaw-pump cycles. After stirring at 60 °C for 17 h, the reaction was terminated by exposure to air. Polymer was precipitated from an excess of diethyl ether three times. After drying overnight in a vacuum oven at room temperature, the light red product was obtained with a monomer conversion of 32.2% from $^1$H-NMR.

## 3. Preparation of CP2VP

After dissolving linear precursors (50 mg) in acetonitrile (1000 mL), the solution was stirred under UV light irradiation for 9 h at room temperature. Pure cyclic polymers were conveniently collected by evaporation of solvent. Repeat the procedure until enough cyclic P2VP were obtained for quaternization.

## 4. Preparation of CPVBIm

Follow similar procedures in "preparation of cyclic P2VP", but with less linear precursors (20 mg) in acetonitrile (1000 mL).

## 5. Preparation of CP2VP-Br

A solution of cyclic P2VP (200 mg) in bromoacetonitrile (5 ml) was stirred at 60 °C for 3 days. After the reaction, the solution was precipitated in diethyl ether (50 ml), and then concentrated and the precipitations were collected. The precipitations were re-dissolved in water (1 ml), and precipitate in THF (20 ml) again. After 3 dissolving-precipitation circle, the CP2VP-Br was obtained upon vacuum dry.

## 6. Preparation of CPVBIm-Br

A mixture solution of cyclic P2VP (200 mg) with bromoacetonitrile (3 ml) in NMP (2 ml) was stirred at 60 °C for 1 days. After the reaction, the solution was dropped into diethyl ether (50 ml), and then concentrated and the precipitations were collected. The precipitations were re-dissolved in water (1 ml), and precipitate in THF (20 ml) again. After 3 dissolving-precipitation circle, the CPVBIm-Br was obtained upon vacuum dry.

## 7. Preparation of Various Cyclic PILs

Ten different cyclic PILs were prepared following identical counter-ion exchange procedures, with molar ratios of 2:1 (salts : PILs). The dissolved PILs solutions (*e.g.* 50 mg CP2VP-Br/CPVBIm-Br in 2 mL water) were dropped into salt solutions (*e.g.* 36 mg NaBF$_4$ in 2mL water), and after filtration, the precipitations were washed 3 times with water, and dried in vacuum for next steps.

## 8. Calculation of counter-ion exchange efficiency

According to the definition of the efficiency of counter-ion exchange, the formula can be shown as (1):

$$E\% = \frac{exchanged\ ionic\ groups}{ionic\ pyridine\ or\ imidazole\ units} \times 100\% \quad (1)$$

### 8.1. The efficiency of counter-ion exchange for CP2VP-Tf$_2$N and CPVBIm-Tf$_2$N by elemental analysis.

Because the efficiency of quaternization of CP2VP-Br was 60%, name is, only 60% pyridine ring unit on the polymer have been ionized. The CP2VP-Tf$_2$N structure was shown in Figure S1. The formula to calculate the content of sulfur was shown in (2), which 64 is the mass of sulfur, 105 is the mass of pyridine ring, 225 is the mass of unexchanged ionic group, 425 is the mass of exchanged ionic group, and x was 0.4, and y plus z was 0.6, E% was z/0.6. The results of elemental analysis was 10.43%, So the efficiency of counter-ion exchange for CP2VP-Tf$_2$N is 71.3%.

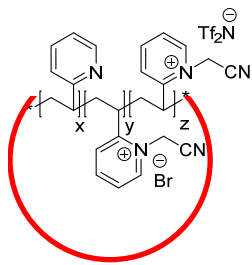

**Figure S1**. The structure of CP2VP-Tf$_2$N.

$$\begin{aligned}\text{Content of Sulfur} &= \frac{z*64}{x*105 + y*225 + z*425} \times 100\% \\ &= \frac{E*0.6*64}{0.4*105 + (0.6 - 0.6*E)*225 + E*0.6*425} \times 100\% \end{aligned} \quad (2)$$

Similarly, because the efficiency of quaternization of CPVBIm-Br was 94%, *i.e.*, only 94% imidazole unit along the polymer backbone have been ionized. The CPVBIm-Tf$_2$N structure was shown in Figure S2. The formula to calculate the content of Sulfur was shown in (3), where 64 is the mass of sulfur, 184 is mass of the neutral group, 304 is the mass of unexchanged ionic group, 504 is the mass of the exchanged ionic group, and x was 0.06, and y plus z was 0.94, E% was z/0.94. The result of elemental analysis was 10.51%, so the efficiency of counter-ion exchange for CPVBIm-Tf$_2$N is 76.2%.

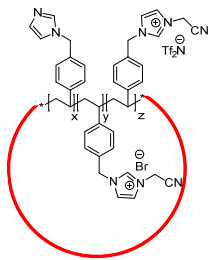

**Figure S2** The structure of CPVBIm-Tf$_2$N.

Content of Sulfur
$$= \frac{z*64}{x*184+y*304+z*504} \times 100\% \quad (3)$$
$$= \frac{E*0.94*64}{0.06*184+(0.94-0.94*E)*304+E*0.94*504} \times 100\%$$

**8.2. The efficiency of counter-ion exchange for CP2VP-BF$_4$ and CPVBIm-BF$_4$ by the ICP-MS.**

Similarly, the efficiency of counter-ion exchange for CP2VP-BF$_4$ and CPVBIm-BF$_4$ could be calculated by the content of boron, which was determined by ICP-MS. The calculation process was the same like E% calculated by content of sulfur for CP2VP-Tf$_2$N and CPVBIm-Tf$_2$N.

For CP2VP-BF$_4$, formula is (4), and the results of ICP-MS was 2.978%, so the efficiency of counter-ion exchange for CP2VP-BF$_4$ is 81.4%.

Content of Boron
$$= \frac{z*11}{x*105+y*225+z*232} \times 100\% \quad (4)$$
$$= \frac{E*0.6*11}{0.4*105+(0.6-0.6*E)*225+E*0.6*232} \times 100\%$$

For CPVBIm-BF$_4$, formula is (5), and the results of ICP-MS was 3.187%, so the efficiency of counter-ion exchange for CPVBIm-BF$_4$ is 93.4%.

Content of Boron
$$= \frac{z*11}{x*184+y*304+z*311} \times 100\% \quad (4)$$
$$= \frac{E*0.94*11}{0.06*184+(0.94-0.94*E)*304+E*0.94*311} \times 100\%$$

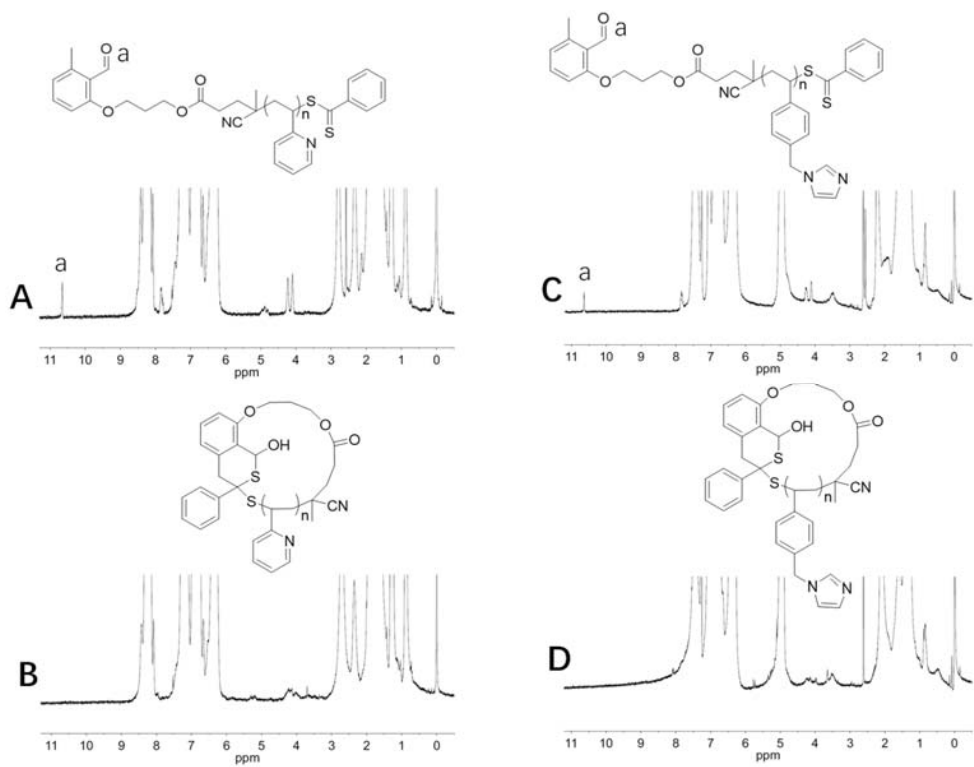

**Figure S3** $^1$H NMR spectra of linear P2VP (A), cyclic P2VP (B), linear PVBIm (C) and cyclic PVBIm (D) in CDCl$_3$.

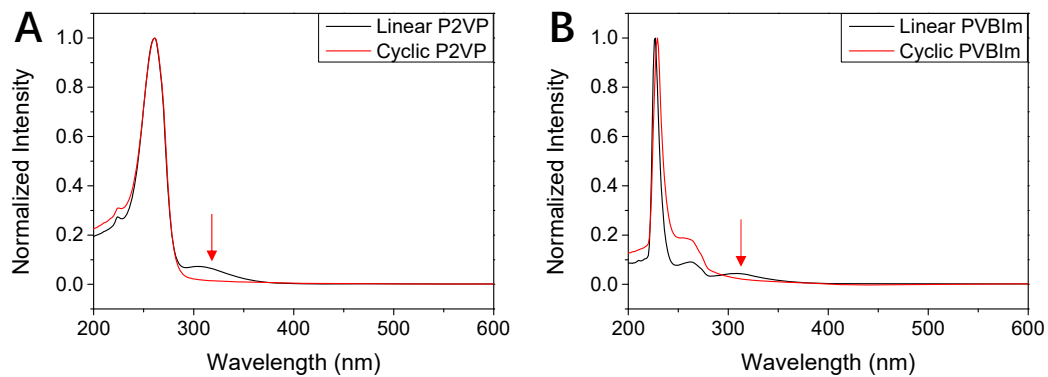

**Figure S4** UV-vis spectra of (A) linear P2VP (black) and the resultant cyclic P2VP (red) in DCM, (B) linear PVBIm (black) and the resultant cyclic PVBIm (red) in DCM.

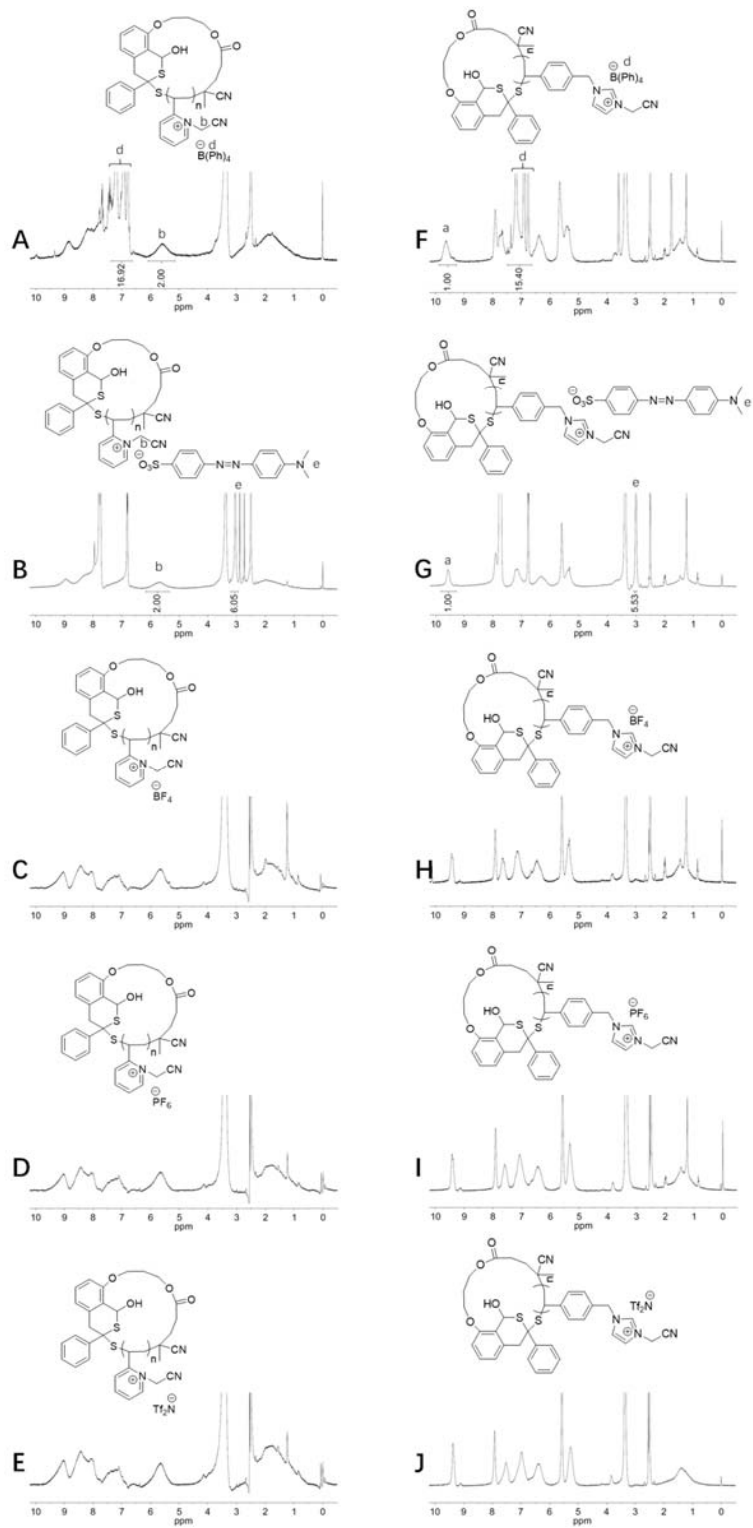

**Figure S5** $^1$H NMR spectra of CPILs: A) CP2VP-B(Ph)$_4$; B) CP2VP-MO; C) CP2VP-BF$_4$; D) CP2VP-PF$_6$; E) CP2VP-Tf$_2$N; F) CPVBIm-B(Ph)$_4$; G) CPVBIm-MO; H) CPVBIm-BF$_4$; I) CPVBIm-PF$_6$; J) CPVBIm-Tf$_2$N; in DMSO-$d_6$.

**Table S1** The content of Nitrogen, Carbon, Hydrogen, Sulfur of CP2VP-Tf$_2$N and CPVBIm-Tf$_2$N.

| SAMPLE | DATA | | | |
|---|---|---|---|---|
| | N /% | C /% | H /% | S /% |
| CP2VP-Tf$_2$N | 7.80 | 43.56 | 3.67 | **10.99** |
| | 7.81 | 43.66 | 3.57 | **10.60** |
| | 7.80 | 43.55 | 3.48 | **10.26** |
| CPVBIm-Tf$_2$N | 10.58 | 37.73 | 2.35 | **11.02** |
| | 10.57 | 37.36 | 2.27 | **10.42** |
| | 10.67 | 37.35 | 2.22 | **10.60** |

**Table S2** The content of Boron of CP2VP-BF$_4$ and CPVBIm-BF$_4$.

| Order number | CP2VP-BF$_4$ | CPVBIm-BF$_4$ |
|---|---|---|
| Sample(g) | 0.0455 | 0.0099 |
| Volume of solution(mL) | 50 | 50 |
| Concentration of solution (mg/L) | 2.710 | 6.31 |
| Dilution rate | 10 | 1 |
| Concentration of Boron of sample (%) | **2.978** | **3.187** |

1. Xie, Y.; Sun, Q.; Fu, Y.; Song, L.; Liang, J.; Xu, X.; Wang, H.; Li, J.; Tu, S.; Lu, X.; Li, J., *J. Mater. Chem. A* **2017,** *5*, 25594.
2. Tang, Q.; Wu, Y.; Sun, P.; Chen, Y.; Zhang, K., *Macromolecules* **2014**, *47*, 3775.